# Cluster-based Position Tracking of Mobile Sensors

Vikram Kumar, Neil W. Bergmann, Izanoordina Ahmad
School of Information Tech. & Elec. Engineering
University of Queensland
Australia

Raja Jurdak, Branislav Kusy
Data61
CSIRO
Pullenvale, Australia

*Abstract*— Tracking movement of mobile nodes has received significant scientific and commercial interest, but long term tracking of resource-constrained mobile nodes remains challenging due to the high energy consumption of satellite receivers. Cooperative position tracking has been proposed for energy efficiency, however, all the cooperative schemes use opportunistic cooperation and optimize for either energy or accuracy. Considering the existence of a reasonably stable group of mobile nodes like animals, birds, and mobile assets, we propose a cluster-based cooperative tracking algorithm, where cluster head centrally coordinates resource usage among cluster members. Variants of this strategy include the use of a cooperative Kalman filter with and without inertial sensor inputs to estimate nodes' positions. We use the Boid flocking algorithm to generate group position movements in 3D and perform various experiments to compare the energy and position accuracy tradeoff of our proposed scheme with individual-based tracking and existing cooperative schemes. We perform the experiments for fixed periodic GPS sampling and dynamic GPS sampling triggered by node position error uncertainty tolerance limit. Experiments results show that in periodic sampling scheme cooperative tracking yields more than one-quarter reduction in energy consumption and more than one-third improvement in position accuracy over individual-based tracking, however, results for dynamic sampling scheme are comparable with existing cooperative scheme.

## I. Introduction and Motivation

Tracking the movement of animals, vehicles and people is useful both for scientific purposes such as finding the range of animal movement, and for logistical purposes, such as finding the closest taxi to service a customer. The Global Positioning System (GPS) has revolutionized location-based services in the past two decades with an accuracy range of 1-50m with consumer grade GPS receivers [1] in outdoor environments. However, GPS is a relatively power-hungry technology, and continuous tracking using GPS is not realistic in resource constrained environments like mobile phones [2][3].

Cooperative positioning is an emerging trend in energy efficient tracking and component duty cycling. Jurdak et al. in [4] implemented group-based sensor duty cycling to reduce the energy consumption. Significant research is currently being undertaken to improve GPS energy consumption and accuracy [5]–[8] using various means of component duty cycling either standalone [9] or group based. However, cooperative positioning efforts in the literature treat group members as individual entities and use opportunistic rather than planned cooperation. Planned centralized cooperation becomes more crucial in dynamic scenarios of resource availability and tracking performance requirements, such as tracking a group of nodes with unpredictable track length using the energy harvesting capable device.

In literature, substantial effort is dedicated to centralized or cluster based schemes focusing on packet routing efficiency, memory efficiency, or other network performance related parameters [10]–[12]. However, to the best of our knowledge, there is no work considering the GPS sensing load distribution using centralized or cluster head based approaches. This paper investigates improvements in energy and localization accuracy of a group of mobile nodes, through sharing the GPS sensing load amongst the group members using a centralized task distribution algorithm.

This work is motivated by several use case scenarios. It is applicable to tracking of mobile unpowered assets, such as inventory, machinery, or reusable containers, in large outdoor areas such as construction or manufacturing sites or during transit and distribution. Additionally, our algorithm can be used to improve tracking of people such as on bus or on bicycles, cooperative schemes can also be helpful in wildlife tracking such as peregrine, whimbrel, dingo, cows and zebras [4], [13]–[16].

Specifically, we explore several fundamental questions regarding the proposed cluster based cooperative tracking:

1) What is the individual nodes' energy–accuracy trade-off of distributing the GPS sampling task among the group of co-located cooperating mobile nodes?

2) Can the energy–accuracy trade-off the tracking be further improved through time-series analysis and modeling, such as using the Kalman filter (KF)?

3) What is the effect on energy-accuracy trade-off of using the individual nodes' Accelerometer & Magnetometer (Acc&Mag) sensors information in centralized cooperative tracking algorithm?

These questions are evaluated for fixed periodic and dynamic GPS sampling scheme.

## II. RELATED WORK

Lee et al. [7] proposed CoMon, a novel cooperative ambiance monitoring platform, which addresses the energy problem through opportunistic cooperation among nearby mobile users using Bluetooth. CoMon uses contact stability profile characteristics, differentiation between acquaintances and strangers, and profit loss analysis for cooperative activity. This work concentrates on general resource sharing among devices whereas we are concerned only about group based energy efficient localization by sharing the motion sensors such as GPS and IMU. A complex negotiation process used in CoMon also makes it unsuitable for resource constrained mobile nodes.

Vukadinovic and Mangold [17] explored the concept of collaborative GPS duty cycling using Wi-Fi ad-hoc connectivity while ensuring application specific error bounds. Whenever a node approaches the error bound limit, first it will request a better position estimate from its neighborhood and wait for a response within a certain time. In the case of a fruitful response, the node will update its position estimate otherwise, the node will go for its own GPS lock followed by a broadcast of new GPS position coordinates. Based upon the analysis of real data traces of visitors in the EPCOT theme park in Florida they reported energy gains from collaboration. However, they assumed the availability of groups and did not consider the costs associated with group searching and group characteristics in addition to their use of energy expensive Wi-Fi in pre and post GPS lock phases.

Jurdak et al. [4], [5] introduced radio contact logging and group-based GPS duty cycling as energy-efficient localization schemes. The concept was tested on cow movement data (GPS and radio range) collected from a real world deployment. They used short range radio communication as a measure of reducing the position uncertainty with the help of neighbours having better position estimates. Their focus was on augmenting GPS location with more energy-efficient location sensors to bound position estimate uncertainty while GPS is off. They also introduced the duty cycling for motion sensors based upon movement patterns and evaluated its impact on empirical data traces of cattle. Here, we build on this concept with centralized sensor sampling strategies to maximize the benefit of neighborhood-based information.

Parker and Valaee [18], [19] focused on measuring the accurate relative distance measurements to avoid vehicle collisions with the use of RSSI-based distance estimates, GPS reading, and vehicle grade IMU. They shared the RSSI based distance and velocity measurements among all the nodes in the group. Each node computed its own motion state based on an Extended Kalman Filter (EKF) for removing the Gaussian errors of the RSSI measurement. Mohammadabadi and Valaee [20] proposed a new distance ranging method based on positive orthogonal codes (POC) and a semi-extended Kalman filter in which the GPS measurement is treated as linear part and neighborhood distance part as a nonlinear part in the same Kalman filter. These works focus on the use of ranging information, IMU information, road map information, and GPS information in order to refine the relative position of the neighborhood nodes in energy-rich environments with vehicle grade IMU information. Our work is focused on the energy and position estimation improvement in a resource constrained environment for a group of uncontrolled mobile nodes using MEMS-based components.

Recently Taniuchi et al.[21] focused on improving the Wi-Fi positioning in indoor environments by using the RSSI-based distances among neighbors. They calculated confidence scores as weights in deciding the position of the neighbour. The confidence of the Wi-Fi-based position is a function of the standard deviation of the multiple Wi-Fi scans for the same point. A lower standard deviation means higher confidence on position and vice - versa. Similarly, the confidence of the Bluetooth is assigned by RSS modeling in the different settings of the indoor environment. Lastly, they used game theory to determine the final position of the nodes.

These works show that performance (energy or position accuracy) gain in tracking is possible when a group of mobile nodes cooperate, for example, by sharing their position estimates and motion characteristics. From the application perspective, in most of the projects, the cooperative approach has been shown useful for the purpose of position error tolerant location-based services (LBS), tracking services etc. However, in such cases, the radius of the group is generally comparable to the position error tolerance limit, which highlights the potential to treat a group of nodes as a single virtual entity rather than multiple individual nodes. Providing a cluster head node with the group information and the ability to set schedules for the entire stable group provides opportunities for improved performance. The focus of this paper is on the investigation of the energy accuracy trade-off offered by the proposed cluster-based cooperative tracking algorithm. Variants of this algorithm include the use of a cooperative KF with and without inertial sensor inputs to estimate nodes' positions.

## III. CLUSTER-BASED COOPERATIVE TRACKING ALGORITHM

In this section, we describe several cooperative tracking algorithms. We will study performance of these algorithms later in the paper, to better understand the key trade-offs that the different variants offer.

### A. Standard Version

Initially, all nodes get a GPS lock, form clusters based on their location (e.g., cluster radius is less than 50m) and start the cluster head election process.

Cluster Head (CH) is selected in an ad-hoc fashion. Nodes independently run random timers and the node whose timer fires first becomes the cluster head, where a tie is resolved by lowest Node ID. In Fig. 3, we simulate the CH selection algorithm over a period of time. Multiple clusters may occur at a particular time depending on the positions of the nodes. A cluster formed in this process will be partially connected with assured connectivity to the CH. One node can be a member of one cluster only and selection among multiple clusters is made on the basis of the distance to the cluster head.

The CH is responsible for scheduling GPS sampling for all of the clusters' members. The schedule is randomized and takes into consideration the available battery status of individual nodes. CH is also responsible for receiving the position updates and distributing these to member nodes.

If a node misses three consecutive position updates, it leaves the current cluster and look for a new cluster if. A departing node initiates a new cluster formation process if no other cluster is in range.

In the absence of new GPS position updates either from the local GPS or from neighbors, the node uses its last position as the current position estimate.

### B. Variant - Cooperative Kalman Filter(CKF)

The basic framework of this version is inherited from the standard cluster based cooperative version defined above. The difference is as given below:

- Each node runs its own KF. Whenever there is a requirement of GPS sampling for KF measurement update, rather than sampling its own GPS, a request to CH is submitted to provide the GPS update.

- Upon the receipt of GPS sampling request from the group member, CH randomly assigns the GPS sampling task among its members, same as in standard version CH is responsible for the distribution of new position value among its members.

- Individual node uses the new position update as measurement update in its KF.

### C. Variant - Cooperative Kalman Filter (CKF) with Acc. &Mag.

The framework is the same as CKF version. In this version each node use its Accelerometer and Magnetometer sensors to update the velocity estimation by KF in between two sampling intervals.

## IV. EXPERIMENTAL SETUP

### A. Movement Model

We first describe a group movement model that we use to generate high temporal resolution spatial data for simulations. Specifically, we use the distributed behavior model [22] for generating birds flock movements proposed by Reynolds. We design a Python-based simulation framework, based on the flocking algorithm, to generate 3D position traces of 20 nodes at the resolution of 1sec in the movement area of 50x50x1 km for one day. Considering that birds/animals are active only 50% of the time (either at night or during the day), we ignore the static position time interval, which may favor our cooperative positioning scheme. Therefore, we generate movements only for 12 hours (43200sec). Nodes from two randomly allocated base camps will start towards a common foraging area. Nodes follow the flocking model and a random motion model during journey and feeding activities, respectively. Furthermore, nodes maintain a speed of approximately 6m/s with internode distance of around 20m. Figs. 1- 5 provide an insight into various aspects of simulated movement data characteristics.

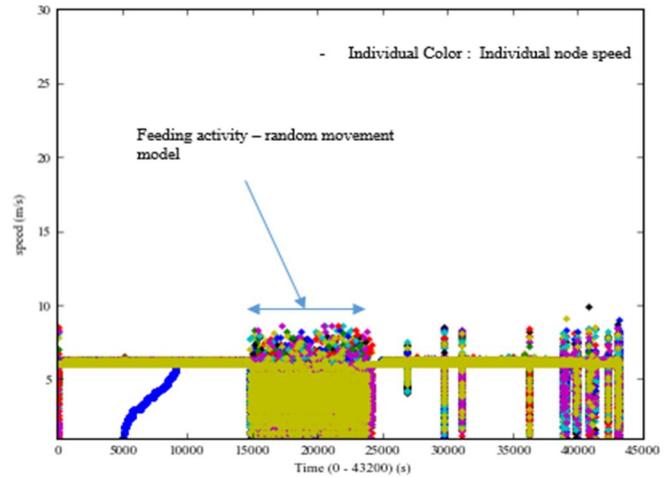
Fig.1. Speed Versus Time - All nodes

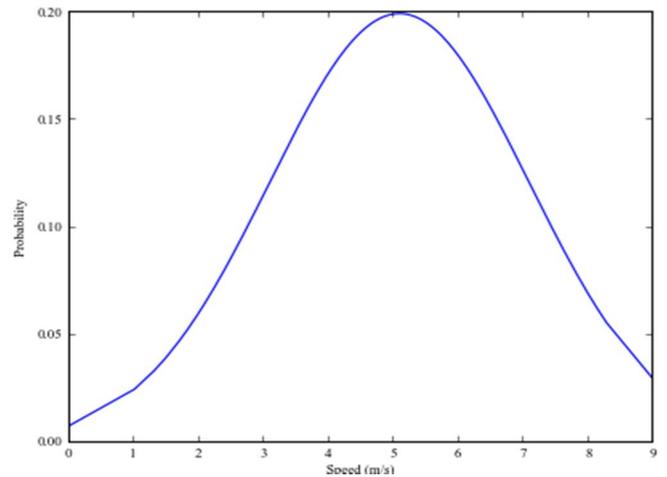
Fig. 2. Node speed distribution

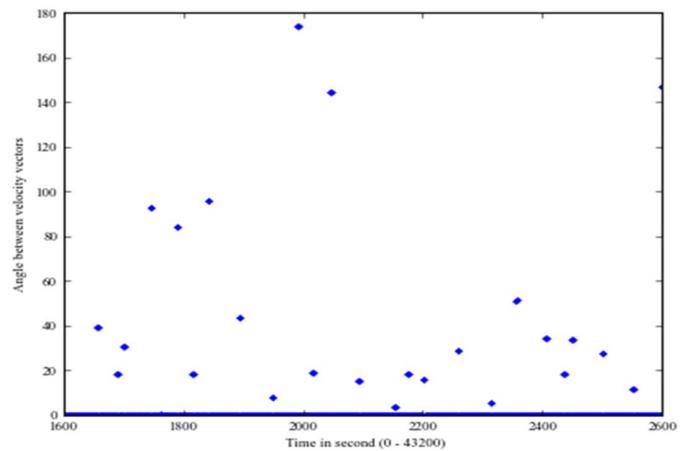
Fig 2a. Time slice view of angle between two velocity vector

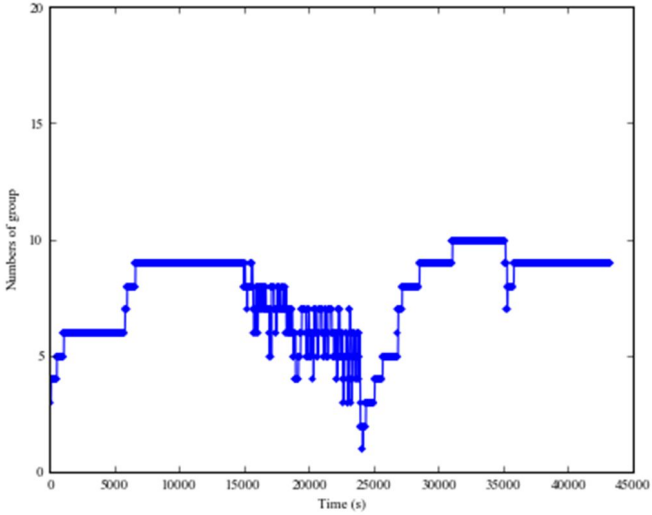

Fig. 3. Variation in group number versus Time

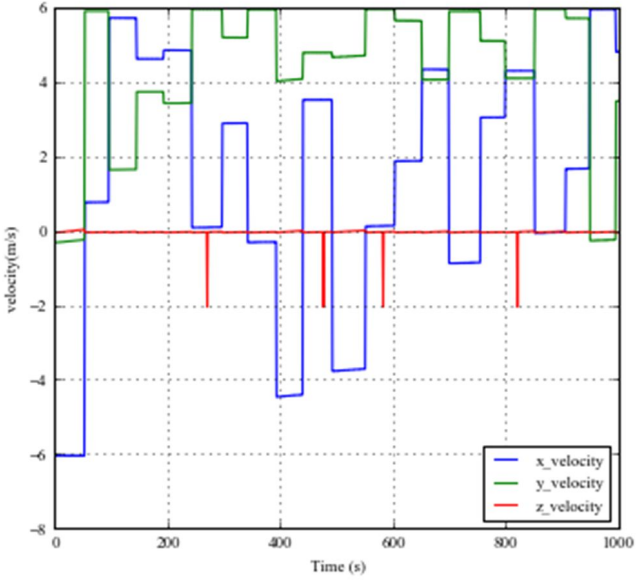

Fig. 4. Time slice View of node 3D velocity

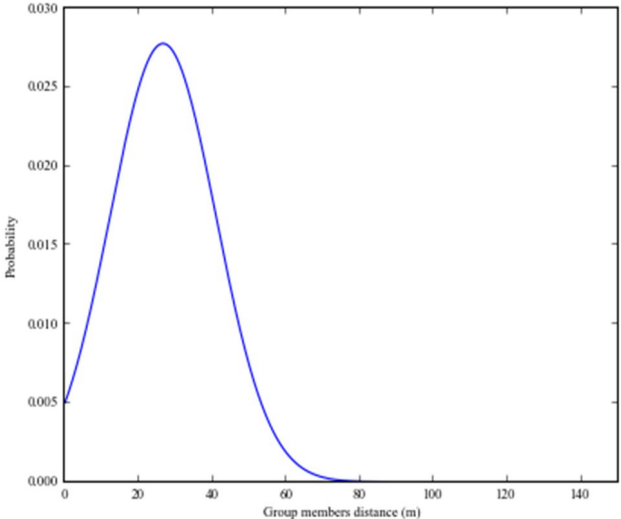

Fig. 5. Group members distance distribution

Table 1. Energy model parameters details

| Component/Activity | Value Used |
|---|---|
| GPS ($P_{GPS}$) | 74mW |
| GPS Lock time for hot start ($T_{Gps}$) (on higher side) | 5s |
| MCU ($P_{MCU}$) | 13.2mW |
| Power consumption radio (Rx+ Tx) ($P_R$) | 99mW |
| Packet Size(S) | 80 bits |
| Channel Bit Rate (CBR) | 256Kbps |
| Packet transmission/Receiving time($T_{RT}$) (S /CBR) | 0.31ms |
| Current Normal mode Acc& Mag ($I_{NM}$) | 110 µA |
| Current Sleep mode Acc& Mag ($I_{SM}$) | 1 µA |
| Power used Acc&Mag in Normal mode ($P_{NM}$) @2.5v | 275 µW |
| Power used Acc&Mag in Sleep Mode ($P_{SM}$) @2.5v | 2.5 µW |
| Time taken for 1 sample, considering maximum sampling rate supported by Acc&Mag is 200Hz (minimum of two) | 5ms |
| Sampling Rate (Acc) | 50Hz |
| Acc& Mag Duty Cycling time ($DC_{Acc\&Mag}$) | 0.25s |
| Platform standby power (Psb) | 1.25 µW |
| Miscellaneous energy compensation ($E_{Mislenergy}$) | 54 J |
| Total simulation period (T) | 43200s |

### B. Energy Model

Energy parameters of this paper are based upon a CSIRO multimodal mobile sensing platform called Camazotz [23]. Camazotz uses a CC430 system on chip (SoC) with GPS, inertial, acoustic, air pressure and temperature sensors, two solar panels, 300mAh Li-Ion battery operating at 3.7V( 3996J), with a total weight just under 30 g targeted at tracking small wildlife such as flying foxes.

We calculate the individual activity (e.g. GPS, Radio, MCU etc.) energy consumption of Camazotz based upon the energy usage reported in [23] and corresponding data sheets of components used. We use a simplified energy model in the simulations: we assume that GPS does not require a cold start considering our sampling interval varies only from 10s to 100s. We assume the energy consumption of radio transmission and reception is the same, and the MCU energy used to collect and process accelerometer and magnetometer samples is negligible given the low sensor sampling rates and low computational complexity of our algorithms. This approach helps us to avoid cycle-accurate estimation of energy consumed by MCU and sensors. Instead, we estimate the total energy nodes require for data processing and standby as miscellaneous energy and focus our analysis on the activities that dominate energy usage: sensor sampling and radio communications. The energy cost of each activity is shown in Table 2 and equations (1) – (4) are summarized in Table 1.

Table 2. Energy consumption of different activities

| Activity/ Component | Total Energy Consumed/Per activity |
|---|---|
| GPS activity ($E_{TotalGPS}$) | 0.436J |
| Radio activity ($E_{TotalTx}$) | 34.78µJ |
| Acc&Mag activity ($E_{Acc\&Mag}$) | 70.62µJ |

$$E_{TotalGPS} = T_{Gps}(P_{GPS} + P_{MCU}) \quad (1)$$

$$E_{TotalTx} = E_{TotalRx} = T_{RT}(P_{MCU} + P_R) \quad (2)$$

$$E_{Acc\&Mag} = DC_{Acc\&Mag}(P_{NM})(1 - DC_{Acc\&Mag})P_{SM} \quad (3)$$

$$E_{Mislenergy} = T * P_{sb} \quad (4)$$

*C. Experiment Details*

We divide the experiments into two different categories based on how the GPS sampling is triggered. The first category is "Fixed periodic sampling", where GPS sampling is triggered at fixed time interval. To the best of our knowledge, there is no existing work in this category, therefore, we compare the performance of the proposed cluster based cooperative strategy and its variants with individual-based sampling. In the individual sampling scheme, each node independently sample and record its GPS position. In the second category "Dynamic sampling", GPS sampling is triggered dynamically based on predefined application specific position error uncertainty tolerance limit. In this category, we compare the performance of our proposed algorithm with existing cooperative scheme proposed by Vukadinovic and Mangold[17].

The goal of our first experiment in "Fixed periodic sampling" category is to understand the energy and accuracy performance of the individual and proposed cluster-based cooperative schemes and how this performance changes at different position sampling intervals varying from 10s to 100s. We define a group size/cluster size of 50m with one hope communication model only. We add Gaussian Noise (0, 10m) to the original GPS position data, which matches GPS performance in low-performance GPS areas such as dense forest, city or cloudy sky etc. based upon the GPS accuracy results presented in [24]. Then we sample the nodes' positions using individual and standard cluster based cooperative sampling strategy defined in section III.

In the second experiment, we apply the CKF variant of cluster based cooperative algorithm defined in section III to track the positions. As described in this scheme we distribute the load of GPS sampling required for the measurement matrix of the KF among the group members. Nodes will share the GPS sampling load as per the cooperative sampling scheme. The position between sampling intervals is computed using linear interpolation to match the full track length of 43200s. We use the 3D velocity parameters provided by GPS as control inputs to the KF and speed accuracy parameter (Sacc) to calculate the system noise covariance matrix (Q). GPS (Pacc), group size and sampling interval are used as an indicators for measurement vector Noise covariance matrix (R). Other details of the CKF are as below:

$$\text{State Matrix} = [P_x, P_y, P_z] \quad (5)$$

$$\text{Control Input} = [V_x, V_y, V_z] \quad (6)$$

$$\text{Measurement Vector} = [Cop_{GPS\_X}, Cop_{GPS\_y}, Cop_{GPS\_z}] \quad (7)$$

$$A = \begin{bmatrix} 1 & 0 & 0 \\ 0 & 1 & 0 \\ 0 & 0 & 1 \end{bmatrix}, B = \begin{bmatrix} dt & 0 & 0 \\ 0 & dt & 0 \\ 0 & 0 & dt \end{bmatrix}, C = \begin{bmatrix} 1 & 0 & 0 \\ 0 & 1 & 0 \\ 0 & 0 & 1 \end{bmatrix} \quad (8)$$

$$Q = \begin{bmatrix} std\_vx^2 * dt^2 & 0 & 0 \\ 0 & std\_vy^2 * dt^2 & 0 \\ 0 & 0 & std\_vz^2 * dt^2 \end{bmatrix} \quad (9)$$

$$R = \begin{bmatrix} std_{Cop_{GPS_X}}^2 & 0 & 0 \\ 0 & std_{Cop_{GPS_Y}}^2 & 0 \\ 0 & 0 & std_{Cop_{GPS_Z}}^2 \end{bmatrix} \quad (10)$$

$P_x$, $P_y$, $P_z$ represent node positions on respective axes. $Cop_{Gps\_x}$, $Cop_{Gps\_y}$, $Cop_{Gps\_z}$, $V_x$, $V_y$, $V_z$ represent the position and velocity on respective axes reported in cooperative GPS activity, dt (1sec) represent time interval for KF update. std_vx, std_vy, std_vz is the standard deviation in the velocity reported by GPS lock on respective axes. std_Cop_GPS_x, std_Cop_GPS_y, std_Cop_GPS_z represents standard deviation in the position, on the respective axes, received through cooperative GPS activity.

The objective of the third experiment was to test whether the availability of accelerometer and magnetometer on every node can help to improve the energy accuracy tradeoff of individual versus cluster-based cooperative scheme. Therefore, we use the "accelerometer and magnetometer" variant of cluster based cooperative scheme. Here we simulated that each node will run its own accelerometer and magnetometer. We added the Gaussian noise value calculated using (11) based upon the parameters given in datasheet.

$$Noise = Noise\ Density * \sqrt{Bandwidth * 1.6} \quad (11)$$

Where bandwidth is the accelerometer bandwidth (50Hz) and noise density (200μg/$\sqrt{Hz}$) is the square root of the power spectral density. The acceleration value calculated after performing all the required processing and conversion of the raw acceleration and magnetometer data used to update the velocity information provided by the cooperative GPS samples. We expected that this will provide a better control input (velocity) to KF, as the cooperative Kalman filter version in the second experiment does not have information on how the velocity is varying in-between sampling intervals.

The same experiments are also performed for the second category "Dynamic sampling ". In this category, we evaluated the performance for position uncertainty limit varying from 50m to 450m. Position uncertainty is assumed to have a linear relationship with velocity and time. For cluster based schemes cluster position uncertainty limit is computed as the average of uncertainty limit of cluster members.

## V. PERFORMANCE EVALUATION

In the experiments, we evaluate the various tracking schemes on the basis of energy consumed and position accuracy. The goal is to identify the best tracking algorithm given some energy budget or accuracy constraints. Each experiment is repeated 20 times to avoid the effects of random variations in a single experiment. The simulation stops either when all the nodes' batteries are exhausted or the track is completed. The results for each category is given below.

### A. Fixed Periodic Sampling

Fig. 6 shows the energy consumption by all the nodes for all the schemes. As expected, cluster-based cooperative schemes outperform the individual scheme in terms of energy consumption. The savings can be as high as 30% over individual sampling for nearly all sampling intervals. The energy efficiency gain comes at the expense of accuracy, as we show next.

Fig.7 compares the mean and standard deviation of the position error of all the nodes for all the schemes. Cluster- based cooperative sampling has higher errors compared to individual sampling rangingg from 42% to 8 % in the sampling interval of 10s to 100s. The use of KF in cooperative position increases the position error performance gap for the higher sampling interval up to 35% in comparison to 8% for cooperation without the KF. This behaviour of the KF can be explained by the node velocity fluctuation on the individual axes within the sampling interval shown in Fig. 4.

Finally, the addition of accelerometer and magnetometer information to the KF (CKF with Acc&Mag) improves the position error performance of the cluster based cooperative tracking. The "CKF with Acc&Mag" scheme shrinks the individual cooperative position error gap range of 42% – 8% to 22% - <1% over various sampling intervals. It is noticeable in Fig. 7 that, at the sampling interval of 100s, the "CKF with Acc&Mag" has the same mean error and a slight reduction in variability in comparison to the individual sampling scheme.

Shifting our focus to the consolidated view of position and battery performance of individual versus cooperative schemes presented in Fig. 8, we can clearly observe benefits as follows. For energy budget of 250J, cluster-based cooperative algorithm performs best with 43% reduction in position error compared to the individual scheme. Alternatively, for a fixed error range of 67 meters "CKF with Acc& Mag" performs better yielding a 27% energy saving compared to the individual scheme.

### B. Dynamic Sampling

The results for position uncertainty limit ranging from 50m to 450m are summarized in Fig. 9. For the initial uncertainty limit of 50m existing cooperative scheme performs better in energy usage with comparable position accuracy performance than proposed cluster-based schemes. The poor performance of cluster-based cooperative scheme can be seen as a result of low position uncertainty limit for a group size of 50m. As described earlier, group position uncertainty limit is computed as the average of position uncertainties of all group members.

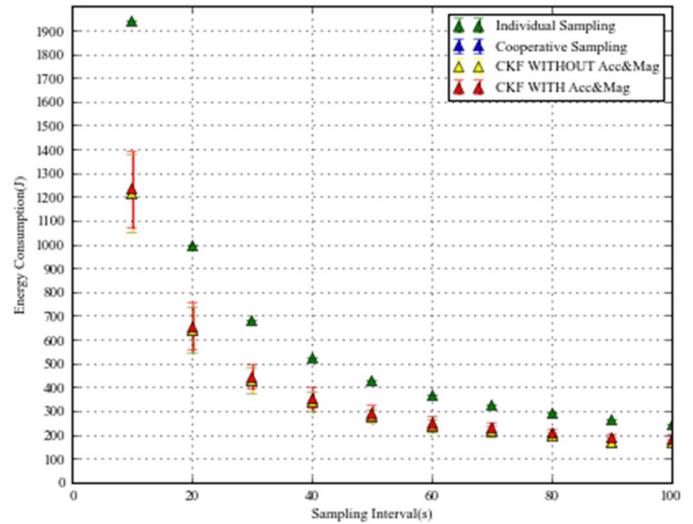

Fig. 6. Mean & Std. (all nodes) energy consumed -periodic sampling

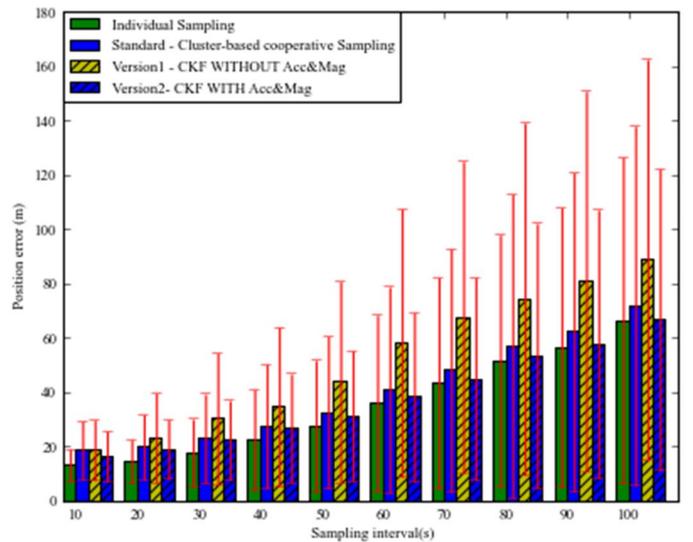

Fig. 7. Position error vs. Sampling interval –periodic sampling

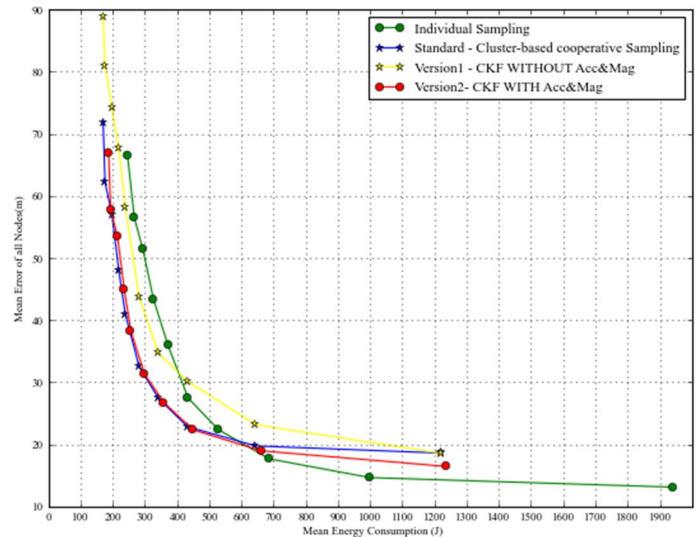

Fig. 8. Mean position error vs. Mean energy for periodic sampling

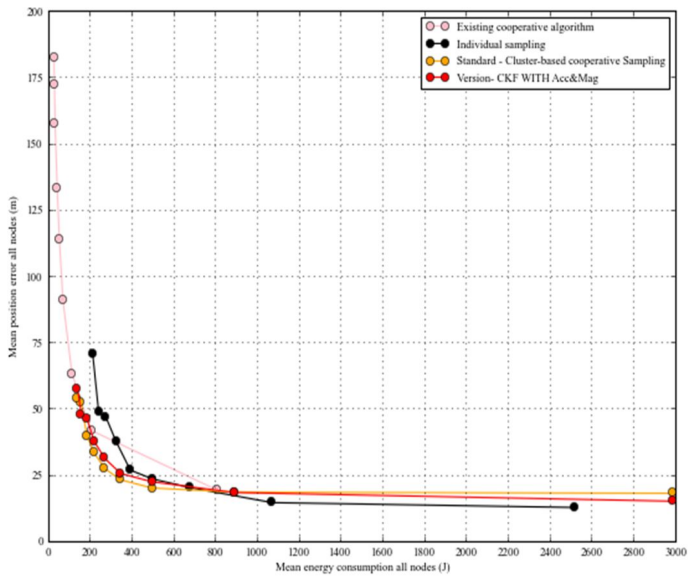

Fig. 9. Mean position error vs. Mean energy for dynamic sampling (Limit varies from 50m to 450m-interval 50m)

For a group radius of 50m, the group position uncertainty limit of 50m can be reached easily, leading to higher GPS sampling rate, causing more energy usage. Overall in this category existing co-operative scheme perform well or comparable to cluster based scheme. However, there are some points like given error range of 22m, cluster based schemes can save energy up to 40% compared to existing cooperative scheme.

Application of the KF in cooperative schemes did not improve much the performance of cluster-based co-operative schemes. This is justified considering that a node in CKFversion gets fairly accurate estimate of velocity only when the node samples GPS. For rest of time, velocity is estimated using noisy inertial sensors or velocity estimates given by neighbouring nodes upon a GPS lock, which in both cases not a good estimate of the current velocity.

## VI. CONCLUSION AND FUTURE WORK

This paper has proposed and evaluated cluster-based cooperative tracking on resource constrained mobile nodes. We have explored three variants of cluster based cooperative tracking, without any KF, with KF, and with KF and inertial sensors. Our results show that the cooperative tracking approach improves energy and accuracy by around one-third over the individual-based sampling strategy in fixed periodic GPS sampling scheme. In dynamic sampling, existing cooperative scheme perform better or comparable to cluster based scheme with only few exception points where cluster base scheme performs better.

Our experiments show that cooperative tracking scheme can give us position error reduction up to 43% for a given energy budget in comparison to the individual based scheme. On the other side "CKF with Acc&Mag" tracking can also save battery life up to 27% while staying within the position error performance range of individual based tracking scheme. For the dynamically triggered GPS sampling category cluster based schemes sometimes can save up to 40% of battery while staying within same error bound.

An interesting direction for future work is to use low power radio interface based RSSI distance estimation for improving the cluster based cooperative positioning estimate. Because the performance of cooperative tracking largely depends on group characteristics such as size, and stability, we will also perform experiments to understand the effects of group characteristics on the energy accuracy trade off. We expect that cluster based cooperative tracking can provide a wide range of energy and accuracy benefits in applications such as wildlife tracking and assets tracking.